%% file: main.tex
\documentclass[conference]{IEEEtran}
\IEEEoverridecommandlockouts

\usepackage{amsmath,amsfonts}
\usepackage{algorithmicx}
\usepackage{algorithm}
\usepackage{array}
\usepackage[caption=false,font=normalsize,labelfont=sf,textfont=sf]{subfig}
\usepackage{textcomp}
\usepackage{stfloats}
\usepackage{url}
\usepackage{verbatim}
\usepackage{graphicx}
\usepackage{cite}
\usepackage{amsmath}

\usepackage{amsmath,amssymb,amsfonts}
\usepackage{graphicx,color}
\usepackage{textcomp}
\usepackage{multirow}
\usepackage{siunitx}
\usepackage{graphicx}
\usepackage{textcomp}
\usepackage{hyperref}
\usepackage{dsfont}
\usepackage{array}
\usepackage{booktabs}
\usepackage{multirow}
\usepackage{comment}
\usepackage{balance}

\graphicspath{{figs/}} 

\usepackage{dsfont}
\usepackage{array}
\usepackage[noend]{algpseudocode}
\usepackage{algorithmicx}

\usepackage{color,soul}

\usepackage{xcolor}

\usepackage{titlesec}

\begin{document}

\title{QR-MO: Q-Routing for Multi-Objective Shortest-Path Computation in 5G-MEC Systems
\thanks{This work has been funded by the Norwegian Research Council through the 5G-MODaNeI project (no. 308909).}
}

\author{
\IEEEauthorblockN{Annisa Sarah}
\IEEEauthorblockA{\textit{Dept. of Electrical Eng. and Computer Science} \\
\textit{University of Stavanger}\\
Stavanger, Norway \\
annisa.sarah@uis.no}
\and
\IEEEauthorblockN{Rosario G. Garroppo}
\IEEEauthorblockA{\textit{Dept. of Information Eng.} \\
\textit{University of Pisa}\\
Pisa, Italy \\
rosario.garroppo@unipi.it}
\and
\IEEEauthorblockN{Gianfranco Nencioni}
\IEEEauthorblockA{\textit{Dept. of Electrical Eng. and Computer Science} \\
\textit{University of Stavanger}\\
Stavanger, Norway \\
gianfranco.nencioni@uis.no}
}



\maketitle

\begin{abstract}
Multi-access edge computing (MEC) is a promising technology that provides low-latency processing capabilities. To optimize the network performance in a MEC system, an efficient routing path between a user and a MEC host is essential. The network performance is characterized by multiple attributes, including packet-loss probability, latency, and jitter. A user service may require a particular combination of such attributes, complicating the shortest-path computation. This paper introduces Q-Routing for Multi-Objective shortest-path computation (QR-MO), which simultaneously optimizes multiple attributes. We compare the QR-MO's solutions with the optimal solutions provided by the Multi-objective Dijkstra Algorithm (MDA). The result shows the favorable potential of QR-MO. After 100 episodes, QR-MO achieves 100\% accuracy in networks with low to moderate average node degrees, regardless of the size, and over 85\% accuracy in networks with high average node degrees.
\end{abstract}

\begin{IEEEkeywords}
multi-objective, shortest path, 5G, MEC.
\end{IEEEkeywords}

\input{01_Introduction}
\input{01-2_RelatedWorks}
\input{02_Problem}

\input{03_Solution}

\input{04_Results}

\input{05_Conclusion}


\bibliographystyle{IEEEtran}
\bibliography{references.bib}

\end{document}

%% file: 01_Introduction.tex
\section{Introduction}
\label{sec:introduction}

\IEEEPARstart{T}{he} integration of 5G-and-Beyond networks and Multi-access Edge Computing (MEC) is an example of integrations between computing and communication, which is also relevant on space-air-ground-sea environment. 5G-MEC systems aim to offer ultra-low latency, high-bandwidth communication, real-time processing, and context-awareness~\cite{Rec_ETSIGR031, 2023_Sarah:5G-MEC-resourceallocation}. A MEC system consists of several computing platforms positioned at the network's edge, which are referred to as MEC Hosts (MEHs). 
In a MEC system, the user traffic should be allocated to the best path to the serving MEH to maintain a high-performance service. Most studies use the shortest path \cite{buzachis2021evaluating,Anwar_trafficsteering}. However, shortest-path algorithms focus on one performance attribute and are not suited to provide heterogeneous services as 5G is meant to do. The network performance can be measured based on different attributes, such as latency, jitter, and packet loss. A user may require different types of services, which may have stringent requirements on one of the attributes rather than another. Therefore, it is important to select a routing path between a user and an MEH, which accounts for multiple cost attributes. This kind of problem is called Multi-objective Shortest-Path (MOSP) problem.

A MOSP problem can be solved by (i) transforming the multiple objectives into a single combined objective. This combined objective is a scalar function that integrates various objectives by assigning weights to each one based on their relative importance~\cite{techpaper:shirdel2018dea}; (ii) generating the whole set of efficient paths as a reference for a decision maker 
\cite{techpaper:MDA_de2021improved}
; (iii) using an approximation method that interacts with the network dynamic condition 
\cite{techpaper:van2014multi}. 

The first approach requires the selection of the weight of each objective before solving the optimization problem. The approach is not flexible because, in the case of a change of the weights, the problem needs to be solved from scratch. The second approach is general and outputs a Pareto set, a set of optimal solutions that are non-dominated to each other and superior to the rest of the solutions. However, these solutions are optimal for a specific network condition, which may not be practical in a dynamic network scenario. The third solution, i.e., the approximation method, uses Genetic Algorithm (GA) 
\cite{ uthayasuriyan2023comparative}
or Reinforcement Learning (RL) \cite{techpaper:van2014multi}. The GA usually needs a complicated representation and suffers in scalability~\cite{uthayasuriyan2023comparative}. 

Regarding the RL-based solutions, Moffaert et al.~\cite{techpaper:van2014multi} addresses general multi-objective optimization, not specifically solving the MOSP problem. In contrast, Yao et al. ~\cite{yao2017efficient} address a specific MOSP problem related to route planning in smart cities, evaluating the efficacy of GA and RL, both of which are approximation methods. Rao~et~al.~\cite{rao2024dar} use RL to solve dynamic MOSP problems, i.e., optimizing delay, packet loss and throughput, by transforming multi-objective to weighted single-objective. Although this approach performs well, it only returns one solution. This solution is dependent on the selected weights assigned to each attribute, rather than representing the Pareto front. The performance of approximating the Pareto front by using RL in a 5G-MEC scenario and its effectiveness compared to optimal solutions remain unexplored.




This paper presented an exploration of the use of RL to obtain an approximation of the Pareto front to solve an MOSP problem in 5G-MEC systems. 
We propose an RL-based approach, called \emph{Q-Routing for Multi-Objective} shortest-path computation (\emph{QR-MO}).
Our initial study investigates how accurate are the solutions found by QR-MO compared to the optimal solutions computed by a traditional deterministic algorithm, such as the Multi-objective Dijkstra Algorithm (MDA). 
The study is performed on multiple 5G-MEH networks in given conditions.

%% file: 01-2_RelatedWorks.tex
\section{Related Works}
\label{sec:relatedworks}



The MOSP problem, which seeks Pareto-optimal paths under multiple objectives, has been addressed through exact methods (e.g., labelling algorithms \cite{raith2009comparison}
and heuristic strategies (e.g., evolutionary algorithms \cite{chang2014greedy}). Raith et al. \cite{raith2009comparison} compare exact algorithms (e.g., label-setting, dynamic programming) and empirically show their poor scalability in dense networks. 
While these methods guarantee optimality, their computational overhead becomes prohibitive in environments like 5G networks, where network conditions might change. This limitation motivates the need for lightweight, adaptive alternatives that can efficiently approximate Pareto-optimal solutions in real-time.

Recent studies demonstrate the use of RL for real-time Pareto set approximation in dynamic networks 
\cite{zhang2023real, khamis2014adaptive}. 
These works provide foundational insights into handling multiple objectives in real-time scenarios, which is relevant in edge computing. However, machine learning methods for multi-objective optimization, including RL, have advanced broadly without sufficient attention to edge computing's unique constraints \cite{zhou2019edge}.

One specific simplified RL approach is called Q-learning. Q-learning has been adapted for routing since Boyan and Littman’s Q-routing framework \cite{techpaper:boyan_1993_QRouting}, which optimizes single-objective paths. However, there is a lack of exploration in using Q-routing for multi-objective path optimization. While some works have applied classical Q-learning to routing problems, they often fail to address the complexities of multi-objective scenarios or leverage improved versions tailored for such tasks \cite{liu2020qmr}. This gap highlights the need for more advanced methods capable of handling multiple objectives efficiently, particularly in 5G-MEC environment.




In 5G-MEC systems, delivering ultra-reliable low-latency communication (URLLC) requires routing solutions that balance multiple objectives, such as latency, jitter, and packet loss, while adapting to network conditions. While single-objective approaches minimize latency \cite{mao2017survey}, it ignore Pareto optimality. Multi-Dijkstra algorithms \cite{techpaper:MDA_de2021improved} compute exact Pareto set but are computationally prohibitive in dynamic 5G networks. RL-based methods such as Q-Routing \cite{techpaper:boyan_1993_QRouting} adapt to network changes but focus on single objectives, leaving a gap in joint optimization of path in 5G-MEC networks.

Our work bridges these gaps with three key aspects:

\begin{itemize}
    \item \textbf{Extension of Q-Routing to Multi-Objective:} while Q-Routing \cite{techpaper:boyan_1993_QRouting} was designed for single-objective optimization, we adapt it to MOSP through heuristic action selection. This is the first application of Q-Routing to multi-objective path selection, to approximate a Pareto set. 
    \item \textbf{Preliminary Validation of the Accuracy:} QR-MO is evaluated by computing the proximity to optimal solutions, i.e. the Pareto set computed by MDA.
    \item \textbf{Context-aware Path Computation in 5G-MEC system:} The evaluation is performed on multiple 5G-MEC networks considering heterogeneous service requirements for latency, jitter, and packet loss, enabling adaptive routing.
 
\end{itemize}



These contributions position QR-MO as an alternative to rigid exact algorithms and single-objective RL, paving the way for real-time, context-aware routing in 5G-MEC systems. 

%% file: 02_Problem.tex
\section{Problem Definition and MDP Representation}
\label{sec:problem}

A MOSP problem can be described as follows.
Given an undirected graph $\mathcal{G} = (\mathcal{V},\mathcal{E})$ with nodes $v \in \mathcal{V}$ and edges $e \in \mathcal{E}$, each edge has $J$ performance attributes, which are also called cost attributes, $\textbf{c}_e = \{c_{e1}, c_{e2}, \dots, c_{eJ}\}$. Solving the MOSP problem means finding a path that optimizes the different cost attributes, which are often conflicting. We have to find a set of strictly non-dominated sets, i.e., the Pareto set. 

Defined $P_{(sn,m)}$ and $Q_{(sn,m)}$ as two paths from the source node $sn$ to the MEH $m$ and defined $c_j(P)$ and $c_j(Q)$ as the $j$-th attribute of the cost vector for paths P and Q respectively,
the path $P_{(sn,m)}$ dominates the path $Q_{(sn,m)}$, denoted as $P_{sn,m} \prec_D Q_{sn,m}$, if the following condition is valid:
\begin{equation}
    \begin{split}
        P_{sn,m} \prec_D Q_{sn,m} \iff \\
        ( c_j(P_{sn,m}) \leq c_j(Q_{sn,m}) \quad \forall j \in [1~..~J]) \\
        \wedge \left( \exists k \in [1~..~J] : c_k(P_{sn,m}) < c_k(Q_{sn,m}) \right)
    \end{split}
\end{equation}
This means that each of the cost attributes of
path $P_{(sn,m)}$ is less than or equal to those of path $Q_{(sn,m)}$. Furthermore, there exists at least one cost attribute path $P_{(sn,m)}$ that
is strictly less than the one of path $Q_{(sn,m)}$. 

Defined \( P_{(sn,m)} \) as a path from the source node \( sn \) to node \( m \). 
\begin{equation}
    \min \mathbf{F}(\mathbf{x}) = \begin{bmatrix}
f_1(\mathbf{x}) \\
f_2(\mathbf{x}) \\
\vdots \\
f_J(\mathbf{x})
\end{bmatrix},
\label{eq:main_objective}
\end{equation}
The objective of the MOSP problem is to minimize all cost attributes associated with the selected path, as expressed in Eq.~\ref{eq:main_objective}, where the cost function for each attribute $j$ is defined as follows.
\begin{equation}
   f_j(\mathbf{x}) = \sum_{e \in \mathcal{E}} c_{ej} \cdot x_e, \quad j = 1, 2, \dots, J. 
   \label{eq:sub_objective}
\end{equation}
In Eq.~\ref{eq:sub_objective}, the function \( f_j(\mathbf{x}) \) calculates the total cost for the \( j \)-th cost attribute across the selected edges in the path. Specifically, for each edge \( e \in \mathcal{E} \), the cost \( c_{ej} \) associated with the \( j \)-th attribute is weighted by the decision variable \( x_e \), which indicates whether the edge \( e \) is included in the path (\( x_e = 1 \)) or not (\( x_e = 0 \)).

The optimization is subject to the following constraints:
\begin{equation}
    x_e \in \{0, 1\}, \quad \forall e \in \mathcal{E}.
    \label{eq:x_cond}
\end{equation}
\begin{equation}
    \sum_{e \in \text{Adj}(v)} x_e =
    \begin{cases}
    1, & \text{if } v = sn, \\
    1, & \text{if } v = m, \\
    2, & \text{if } v \in \mathcal{V} \not\in \{sn, m\}, \\
    0, & \text{otherwise}.
    \end{cases}
    \label{eq:flow_constraint}
\end{equation}
Eq.~\ref{eq:x_cond} ensures that each edge \( e \in \mathcal{E} \) is either selected or not, represented by the binary decision variable \( x_e \).


Eq.~\ref{eq:flow_constraint} shows the flow conservation constraint that ensures the selected edges form a valid path. At the source node $sn$, exactly one edge should be selected for the path. Similarly, at the destination node, exactly one edge should be selected for the path. For any intermediate node, exactly two edges must be selected. Lastly, nodes that are not part of the path have no selected edges.

%% file: 03_Solution.tex
\section{Proposed Solution}
\label{sec:solution}

RL traditionally optimizes a single scalar reward, yet many real-world problems involve multiple conflicting objectives \cite{hayes2022practical}. Multi-objective RL can be categorized as: (i) utility-based vs Pareto-based, and (ii) single-policy vs multi-policy \cite{hayes2022practical}.

Utility-based approaches focus on optimizing a single scalarized reward function, which combines multiple objectives into a weighted sum or other forms of aggregation. Although this simplifies the optimization problem, it requires predefined weights or utility functions, which are often difficult to determine \cite{ruadulescu2020utility,rao2024dar}. On the other hand, Pareto-based approaches aim to identify a set of Pareto-optimal solutions, each representing a trade-off where improving one objective would degrade another. This provides decision-makers with more flexibility in choosing the most appropriate solution based on specific preferences or priorities \cite{mehta2022pareto,perera2023graph,wang2020thirty}.

Single-policy methods aim to learn a single optimal solution, typically tailored to a fixed utility function. However, these methods do not accommodate uncertain preferences, which limits their applicability in real-world scenarios \cite{hayes2022practical}. In contrast, multi-policy approaches generate a diverse set of solutions, offering greater adaptability to varying or evolving preferences, which is crucial in environments with competing objectives or changing constraints \cite{wang2020thirty}.

By leveraging Pareto-based and multi-policy approaches, multi-objective RL can better address the complexities of real-world decision making, providing not only a broader set of solutions, but also more robust adaptability to diverse needs. The proposed QR-MO algorithm builds on these principles by approximating the Pareto front and adopting a multi-policy approach. Using RL and heuristic action selection, QR-MO identifies a diverse set of Pareto-optimal solutions, each representing a trade-off between conflicting objectives. This multi-policy capability ensures that QR-MO adapts effectively to evolving preferences in real-time environments, such as 5G-MEC systems, where conditions and requirements are constantly changing.

\subsection{Q-learning}

\emph{Q-learning} is a model-free algorithm that learns the state-action value (or Q value). The Q-value represents the expected total value for a particular action $a$ in a given state $s$. The Q-value is generally updated according to the update rule in Eq.~\ref{eq:Q_update}. To update the Q-values, we have to sum three factors: the current values $(1 - \alpha) \cdot Q(s,a)$, the reward $r$ of taking action $a$ from state $s$, and the maximum reward that can be obtained from next state $s'$, $\max\limits_{a} Q(s',a')$. The Q-learning algorithm stores the state-action values in a Q-table, which will be updated until they converge or certain criteria are met.
\begin{equation}
Q(s,a) \leftarrow (1 - \alpha) \cdot Q(s,a) + \alpha \cdot ( r + \gamma \cdot \max\limits_{a} Q(s',a'))
\label{eq:Q_update}
\end{equation}
\begin{equation}
    Q(s,a) \leftarrow (1 - \alpha) \cdot Q(s,a) + \alpha \cdot ( \mathbf{c_{s,a}} + \min\limits_{a' \in \text{neighbors of $s'$}} Q(s',a'))
    \label{eq:Q_routing}
\end{equation}


Boyan and Littman~\cite{techpaper:boyan_1993_QRouting} proposed a modified Q-learning for a routing problem, namely Q-routing. The Q-routing does not need the discount factor as in the generic Q-learning technique ($\gamma = 1$). The Q-routing aims to minimize the future cost (i.e., minimize the total latency) instead of maximizing the future reward. For the Q-routing, the Q-update equation will be slightly changed as in Eq.~\ref{eq:Q_routing}. The state $s$ is the current node $v$, where $|S| = |\mathcal{V}|$ and the action $a$ is the edge $e$ from the current node minus the incoming edge. 
The Q-values of state $s$, taking action $a$ that leads to the next state $s'$, are updated by the sum of the current value, the cost attributes $\mathbf{c_{s,a}}$ of the action $a$, and the minimum cost attributes of the neighbors of the next state $s'$ to reach the end node.  

\subsection{QR-MO}

The problem with the classical Q-routing technique is that it can only consider one cost. Algorithm~\ref{alg:ql} shows the modified Q-routing algorithm that addresses the multi-objective problem, namely \emph{Q-Routing for Multi-Objective} problem (\emph{QR-MO}). The Q-routing implementation in our work is a modification of the code from \cite{web:yuan2023_reinforcement}. In contrast, the Q-table in the proposed QR-MO stores multiple values for each $Q(s, a)$, corresponding to the various cost attributes considered. To choose the best action $a$ in each state $s$, the QR-MO simultaneously considers different criteria by using a heuristic approach. Consequently, QR-MO can generate $J$ solutions, i.e., each solution is the best solution for a specific attribute. 


\begin{algorithm}
\small
\caption{QR-MO for Path Selection}
\label{alg:ql}
\begin{algorithmic}[1]

\State \textbf{Initialize:} Load the network graph $\mathcal{G}$; A start node $n^S$ and end node $n^E$; Number of episodes $N$; Cost matrix $R$; Uniform $Q$; Empty memory $\mathcal{B}$; Learning rate $\alpha$; $\epsilon$ for the epsilon-greedy action selection.

\For{$i$ \textbf{to} $N$}
   \State Initialize $t = 0$
   \State Initialize the current state to the start node $s_t = n^S$
   \While{$s_t \neq n^E$}
        \State Check next possible nodes from $s_t$
        \State Choose $a_t$ by using an $\epsilon$-greedy policy as below:
        \State $a_t$ =
          $ \begin{cases}
                \text{random $a$},& \text{if } p = \epsilon \\
                \text{\textcolor{teal}{DominanceSelection}($a_{t-1}, s_t$, $Q_t$)}, & \text{if } p = 1 - \epsilon
            \end{cases}$
        \State Perform the chosen action $a_t$, transition to $s_{t+1}$
        \State $Q(s_t, a_t) \leftarrow$ \textcolor{teal}{UpdateQ}($R$, $Q_t$, $s_t$, $a_t$, $\alpha$)
        \State Update the current state $s_t = s_{t+1}$
    \EndWhile

    \State Get the route $l_i \leftarrow \{n^S, ..., n^E\}$ and Q-values $Q_i$
    \State Check and store the best cost and path $\mathbf{B} \leftarrow $ \textcolor{teal}{UpdateBestPath}($l_i$, $G$,$\mathbf{B}$,$Q_i$,\textit{i})
\EndFor

\State \textbf{Return:} (1) $Q$ (2) $\mathbf{B}$

\end{algorithmic}
\end{algorithm}


Algorithm~\ref{alg:ql} presents the detailed steps for the QR-MO. We first initialize a network graph $G$, set a start node $n^S$ and an end node $n^E$,
a cost matrix $R$, uniform Q-values for all pair of state $s \in S$ and action $a \in A$, and an empty variable $B$ to store the record of best Q-values, best paths and best cost of each cost attribute $j$ while learning throughout episodes. We also set the RL hyper-parameters: learning rate $\alpha$ and epsilon greedy parameter $\epsilon$. Then, for each episode, the QR-MO agent learns the path from starting node $n^S$ to reach end node $n^E$ by selecting a proper action $a$ on the current state $s$. We employ $\epsilon$-greedy policy to select an action, meaning that the QR-MO agent will take random action with probability $\epsilon$ and a greedy action (i.e., take the best action) with probability 1-$\epsilon$. The $\epsilon$-greedy policy is useful to balance the exploration and exploitation of the learning to seek an optimal policy. The best action for the QR-MO can be selected by using a heuristic algorithm called ~\hyperref[proc:domsel]{\textcolor{teal}{DominanceSelection}}. The Algorithm \hyperref[proc:domsel]{\textcolor{teal}{DominanceSelection}} has been adapted from paper \cite{techpaper:MDA_de2021improved} and used to evaluate the domination of the cost attributes of neighbouring edges. 


\begin{algorithm}
\small
\caption{\textcolor{teal}{DominanceSelection}}
\label{proc:domsel}
\begin{algorithmic}[1]
\State \textbf{Input:} Previous selected action $a_{t-1}$; Current state $s_t$; Learned policy $\mathbf{Q}$.
\State \textbf{Initialize:} Defined $\mathcal{A}_{keys}$ as the set of the possible actions (edges to neighboring nodes) from state $s_t$ excluding the previous selected action $a_{t-1}$, i.e., incoming edge; An empty dictionary $D$ to store dominance scores for each action $D(a) = 0 \quad \forall a \in A_{keys}$  
\For{each $(a_x, a_y) \in C_{pairs}$}
    \For{each cost index $j$}
        \If{$Q^j(s_t, a_x) \leq Q^j(s_t, a_y)$} 
         \State   Increment $D(a_x)$ by 1
        \EndIf
        \If{$Q^j(s_t, a_x) > Q^j(s_t, a_y)$} 
         \State   Increment $D(a_y)$ by 1
        \EndIf
    \EndFor
\EndFor

\State \textbf{Select the best action:} 
\State $a_{dom} \gets \arg\max_{a \in A_{keys}} D(a)$

\State \textbf{Return:} $a_{dom}$

\end{algorithmic}
\end{algorithm}

The Algorithm~\hyperref[proc:domsel]{\textcolor{teal}{DominanceSelection}} needs three inputs: previous selected action $a_{t-1}$, current state $s_t$, and Q-values at timestep $t$, $Q_t$. The previously selected action $a_{t-1}$ is the action that made the agent visit the current state $s_t$, i.e., incoming edge. We initialize a dictionary $D$ to count the dominance scores for each action. Then, all possible actions from current state $a \in \mathcal{A}_{keys}$ are categorized as pairs $(a_x,a_y) \in C_{pairs}$ where $x \in \mathcal{A}_{keys}, y \in \mathcal{A}_{keys}, x \neq y $. All possible actions $\mathcal{A}_{keys}$ are all edges connecting to the current node $s_t$, except the incoming edge, which is the previously selected action $a_{t-1}$. 
For each pair ($a_x,a_y$), we investigate the dominance based on each cost attribute $j$ and count the dominance scores of all actions. The best action $a_{dom}$ is the one with the highest scores among all possible actions $\mathcal{A}_{keys}$ and is returned to Algorithm~\ref{alg:ql}. 


\begin{algorithm}
\small
\caption{\textcolor{teal}{UpdateBestPath}}
\label{proc:UpdateBestPath}
\begin{algorithmic}[1]
\State \textbf{Input:} Network graph $\mathcal{G}$; Q-values matrix for episode $i$ $Q_i$; Start node $n^S$ and end node $n^E$; Route of current episode $i$, $l_i$ from $n^S$ to reach end $n^E$, Stored memory $\mathbf{B}$=[$\{l^B_1,\mathbf{c}^{l^B_1},Q^B_1\},\{l^B_2,\mathbf{c}^{l^B_2},Q^B_2\}, \{l^B_3,\mathbf{c}^{l^B_3},Q^B_3\}$] 

\State \textbf{Initialize:} Costs of route $l_i$, $ \mathbf{c}^{l_i} = \{c^{l_i}_1, c^{l_i}_2, c^{l_i}_3\}$ 

\For{each $j$ in $c_j$}
    \If{$c^{l_i}_j$ \textless $c^{l^B_j}$}
        \State Update best cost $j$ on memory $c^{l^B_j}$ $\leftarrow$ $c^{l_i}_j$
        \State Update best route for cost $j$ on memory  $\mathbf{c}^{l^B_j} \leftarrow l_i$
        \State Store $Q^B_j \leftarrow  Q_i$
    \EndIf
\EndFor
\State \textbf{return} updated $\mathcal{B}$

\end{algorithmic}
\end{algorithm}


In Algorithm~\ref{alg:ql}, the Q-value $Q(s_t,a_t)$ is then updated using Eq.~\ref{eq:Q_routing}. After reaching the end node $n^E$, the route $l_i$ and Q-values $Q_i$ are stored. Lastly, we have to check the best cost, route, and Q-values $B$ by using the Algorithm~\hyperref[proc:UpdateBestPath]{\textcolor{teal}{UpdateBestPath}}. The Algorithm~\hyperref[proc:UpdateBestPath]{\textcolor{teal}{UpdateBestPath}} evaluates the learned route, cost, and Q-values on episode $i$ by comparing it with the tuple $\{l^B_j,\mathbf{c}^{l^B_j},Q^B_j\}$, where $l^B_j$ is the current best route of attribute $j$, $\mathbf{c}^{l^B_j}$ is the tuple of all cost attributes concerning the route $l^B_j$, and $Q^B_j$ is the Q-values or policy to generate the route $l^B_j$. The stored memory is then compared. If the learned cost of attribute $j$ on episode $i$ is better than the best cost $j$ in the memory $B$, then the memory is updated. This algorithm ensures that the QR-MO returns multiple solutions considering all cost attributes simultaneously, compared to classic Q-routing, which only returns one solution.

\subsection{Complexity Analysis}










The QR-MO algorithm operates on a graph with \(|\mathcal{V}|\) nodes, \(|\mathcal{E}|\) edges, and \(J\) cost attributes. It iterates over \(N\) episodes, refining paths via a While loop that explores the graph and selects actions at each node. The Algorithm \hyperref[proc:domsel]{\textcolor{teal}{DominanceSelection}} compares action pairs, requiring \(O(d^2 \cdot J)\) operations per node with degree \(d\). Since traversal involves up to \(|\mathcal{V}|\) nodes, the per-episode complexity is \(O(|\mathcal{V}| \cdot J \cdot d^2)\), where \(d \leq |\mathcal{E}| / |\mathcal{V}|\).
The Algorithm \hyperref[proc:UpdateBestPath]{\textcolor{teal}{UpdateBestPath}} evaluates paths with \(O(J)\) operations per path, which is negligible compared to selection.

Thus, the total complexity of QR-MO is $O(N \cdot |\mathcal{V}| \cdot J \cdot d^2)$. Substituting \(d^2 \sim (|\mathcal{E}| / |\mathcal{V})|^2\), it simplifies to $O(N \cdot J \cdot |\mathcal{E}|^2 / |\mathcal{V}|)$. For sparse graphs (\(|\mathcal{E}| \sim |\mathcal{V}|\)), this reduces to $O(N \cdot |\mathcal{V}| \cdot J)$, while for dense graphs (\(|\mathcal{E}| \sim |\mathcal{V}|^2\)), it scales as $O(N \cdot |\mathcal{V}|^3 \cdot J)$.

Comparing QR-MO to existing Pareto optimal algorithms such as MDA highlights key efficiency trade-offs. As presented by Casas et al. \cite{techpaper:MDA_de2021improved}, MDA maintains multiple non-dominated labels per node. Let \(L\) be the average number of labels per node and \(L_{\max}\) be the maximum number of labels at any node, then MDA has a complexity equal to
$O\left(|\mathcal{V}| \cdot J \left( L \cdot \log |\mathcal{E}| + L_{\max}^2 \cdot |\mathcal{E}| \right) \right)$.

Heap operations introduce a logarithmic term, while dominance checks scale quadratically with \(L_{\max}^2\) when \(J \geq 3\), making MDA expensive for high-dimensional cases (i.e., scenarios where the number of cost attributes \(J\) is high). As \(J\) increases, the number of non-dominated solutions grows, making dominance checks more computationally expensive. 
The term \(L_{\max}^2 \cdot |\mathcal{E}|\) dominates for high \(J\) (\(J \geq 3\)). 
For high \(J\), the number of stored labels per node increases due to the growth in non-dominated solutions. The dominance check in MDA requires comparing each label against others, leading to a worst-case complexity of \(O(L_{\max}^2)\) per edge. Since \(L_{\max}\) increases with \(J\), the term \(L_{\max}^2 \cdot |\mathcal{E}|\) dominates, making MDA less efficient in high-dimensional settings.

In practical applications, QR-MO is preferable for sparse graphs or high \(J\), avoiding quadratic label growth that hampers MDA in multi-objective settings. However, MDA remains effective in low-dimensional, highly connected graphs, where efficient queue operations mitigate dominance check overhead.

%% file: 04_Results.tex
\section{Experimental Results}
\label{sec:result}

This section describes the evaluation scenario: network topology, reference solution, evaluation metrics, and other simulation settings. Further, the results and discussions are presented.



We use the network graphs from dataset~\cite{techpaper:xiang2021dataset}, which has three synthetic graphs and one real network scenario for a 5G-MEC system in Milan City Centre (MCC). The MCC graph has been interpolated from OpenCellID. 
There are four network topologies, i.e., 25N50E, 100N150E, 30N35E and 50N50E. 25N50E means that the network consists of 25 nodes and 50 edges. These network topologies have been selected because they represent a 5G-MEC system with a variety of network characteristics. 25N50E and 100N150E have a high average node degree, with 3 or 4 edges per node, respectively. MCC and 50N50E have a low average node degree of 2.3 and 2, respectively. 

The dataset provides only the network structural information and assumes the same cost for all edges. We modify the cost values and consider three cost attributes which are randomly generated by using a uniform distribution: (1) packet-loss probability with a Probability Density Function (PDF) of 1/3 $U(0.0005,0.1)$+ 2/3 $U(0,0.0005))$, (2) latency 1/3 $U(5,10)$ + 2/3 $U(1,5))$ ms, and (3) jitter 1/3 $U(3,5)$ + 2/3 $U(1,3))$. 
The packet loss probability, latency, and jitter values are taken from measurement in a 5G Campus Networks~\cite{Rischke_5Gcampusnetwork}. Rischke~et~al.~\cite{Rischke_5Gcampusnetwork} indicate that the packet loss probability of a single transmission in a testbed of non-standalone 5G, with a packet size of 128 to 256 bytes, ranges between 0 and 0.15.
The latency of the same type of network varies between 1 ms to 10 ms ~\cite{Rischke_5Gcampusnetwork}. For jitter, the values are derived from \cite{Rischke_5Gcampusnetwork} and \cite{Ronteix_reducing_latency_and_jitter}, which show that the jitter varies from 1 ms to half of the maximum latency.

\subsection{Performance Metrics}
\label{subsec:performance_metrics}


In this work, we use the Pareto set computed with MDA as the baseline solution to evaluate the performance of QR-MO. An evaluation metric called \emph{Distance to Pareto Set (DPS)} is computed to determine the proximity of the QR-MO solution to the Pareto set. The DPS consists of the Euclidean distance between the QR-MO solutions and the Pareto set. 
The $i$-th solution of the Pareto set is denoted as $\mathbf{\Psi_i} = \{\Psi_{ij} \quad \forall j \in [1\ ..\ J]\}$, where $\Psi_{ij}$ is the value of the $j$-th cost attribute. There is no predetermined number of solutions in the Pareto set, i.e., the number of $i$-es is unknown.
Instead, QR-MO always returns $K$ solutions, and the $k$-th solution is denoted as $\mathbf{\Omega_k} = \{\Omega_{kj} \quad \forall j \in [1\ ..\ J]\}$. 
Therefore, in QR-MO, the number of solutions is predetermined and is equal to the number of cost attributes $K=J=3$ because each of the QR-MO solutions optimizes one of the attributes. As previously mentioned, the cost attributes considered in this study are packet loss probability, latency, and jitter.

The DPS can be calculated as follows. First, given the different scales of the cost attributes, the value of each attribute is normalized to the range $[0,1]$. For each attribute $j \in [1\ ..\ J]$, the normalization factor $f_j$ is calculated as $f_j = \max_{i,k}\{\Psi_{ij},\Omega_{kj}\}$.
The QR-MO solutions and each solution of the Pareto set are normalized by using the corresponding normalization factor $f_j$. The normalized solutions are obtained as 
$\mathbf{\Omega_k^{norm}} = \{\frac{\Omega_{k1}}{f_1}, \cdots, \frac{\Omega_{kj}}{f_J}\}$ and $\mathbf{\Psi_i^{norm}} = \{\frac{\Psi_{i1}}{f_1}, \cdots, \frac{\Psi_{ij}}{f_J}\}$.  
The distance between two solutions $d_{ki}$ is calculated as $ d_{ki} = \sqrt{ \sum_{j=1}^{J} \left( \Omega_{kj}^{norm} - \Psi_{ij}^{norm} \right)^2 }$. 
Then the DPS is the minimum distance between the two sets of solutions $DPS = \min_{k,i} d_{ki}$.
The lower the $d_{ki}$ value, the closer the QR-MO solutions are to the Pareto set.

Correctness is another metric to evaluate whether one of the $K$ QR-MO solutions is a part of the Pareto set; if it is true, then the correctness equals 1. The \emph{average correctness} is the mean of the correctness of QR-MO solutions throughout all the simulation runs. As the QR-MO returns $K$ path solutions, i.e., one for each cost attribute,
it is important to evaluate how many solutions of the Pareto set the QR-MO can find. The \emph{average number of correct solutions} shows the mean number of QR-MO solutions that are part of the Pareto set.

\subsection{Simulation Setting and Result Discussion}
We aim to compare our QR-MO solution to its optimality. Therefore, we use MDA as a baseline algorithm to evaluate our proposed solution, since it returns optimal solutions. MDA is a label-setting algorithm introduced in the paper \cite{techpaper:MDA_de2021improved} to address the MOSP problem. MDA generates all solutions in the Pareto set, identified through the list of non-dominated labels of the nodes. The concept of non-domination is explained in Section~\ref{sec:problem}. MDA operates under the assumption that the cost attributes of each edge are summable. While this is true for latency and jitter, it does not apply to packet loss probability. To address this limitation, packet loss probability must be converted to logarithmic form to become summable.

The QR-MO uses an $\epsilon$-greedy action selection with $\epsilon = 0.1$ and the learning rate $\alpha = 0.7$. The hyperparameters of QR-MO are decided after the empirical studies, and the selected values give the best overall performance. A QR-MO agent starts exploring the network from the start node to the end node. We set the maximum episode number $N$ 
to 100, and one episode refers to a period when a QR-MO agent has a single exploration from a starting node and reaches an end node. We generate $25$ different instances: $5$ pairs of start and end nodes, and for each pair, we conduct  $5$ simulation runs. On each run, we have a randomization of $\epsilon$ greedy action selection; thus, we show the average and the $95\%$ confidence interval. The experiments are performed on a laptop equipped with 8 virtual CPUs, a 2.8 GHz processor, 32 GB of RAM, and Python 3.9.



\begin{figure*}[!t]
\centering
\subfloat[]{\includegraphics[width=1.75in]{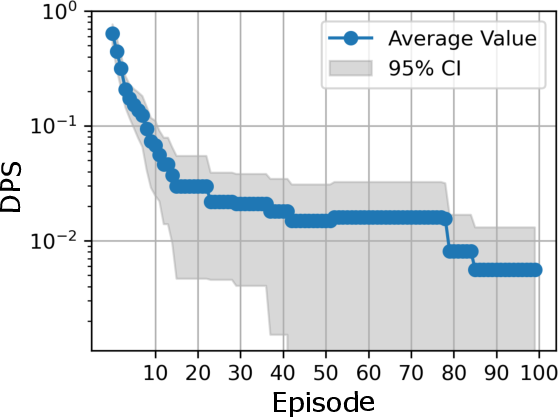}%
\label{fig:DPS-all_a}}
\hfil
\subfloat[]{\includegraphics[width=1.7in]{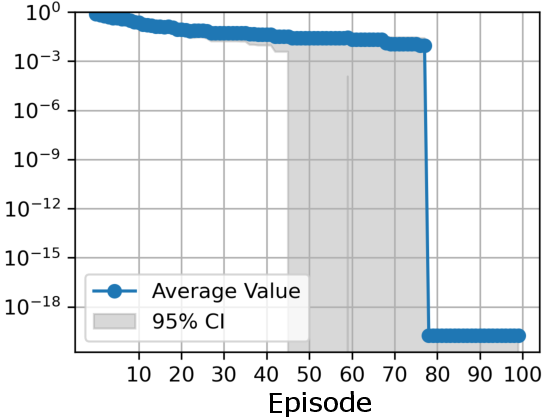}%
\label{fig:DPS-all_b}}
\subfloat[]{\includegraphics[width=1.7in]{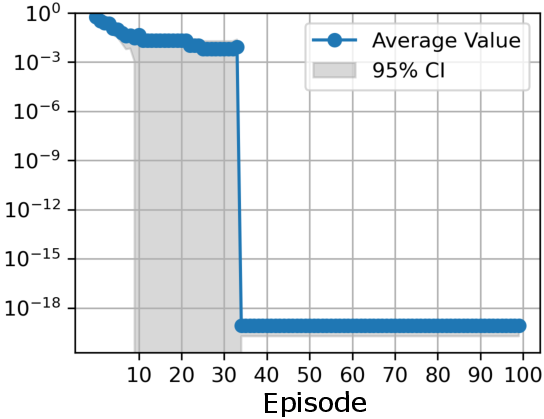}%
\label{fig:DPS-all_c}}
\hfil
\subfloat[]{\includegraphics[width=1.7in]{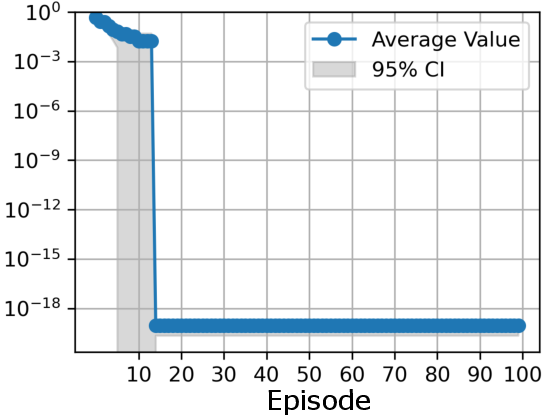}%
\label{fig:DPS-all_d}}
\caption{Average DPS of QR-MO of network (a) 25N50E, (b) 100N150E, (c) MCC (30N35E), and (d) 50N50E}
\label{fig:DPS-all}
\end{figure*}


\begin{figure}[!t]
\centering
\subfloat[]{\includegraphics[width=1.48in]{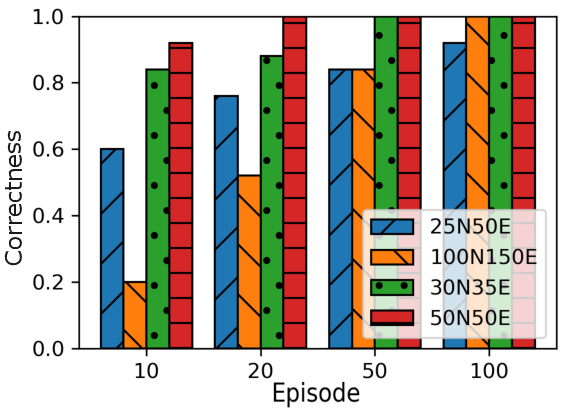}}%
\label{fig:correctness_a}
\hfil
\subfloat[]{\includegraphics[width=1.48in]{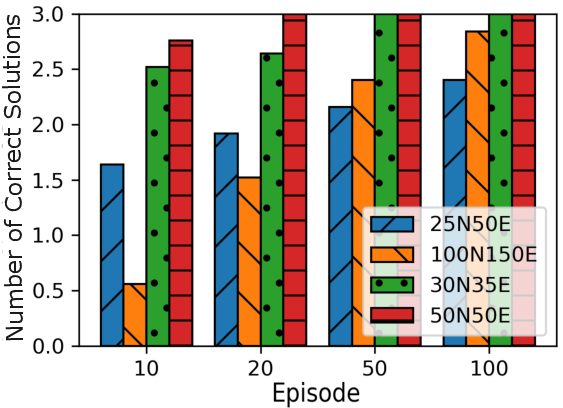}}%
\label{fig:correctness_b}
\caption{Comparison of (a) average correctness of QR-MO and (b) the average number of correct solutions of QR-MO for different topologies}
\label{fig:correctness}
\end{figure}


Fig.~\ref{fig:DPS-all} illustrates the DPS across the various network topologies over $N=100$ episodes. The QR-MO performs exceptionally well for both the MCC and 50N50E. The DPS decreased to nearly 0, with around 30 episodes on the MCC network and around 10 episodes on the 50N50E network. MCC and 50N50E are small networks with low node degrees of 2.3 and 2.0 edges per node, resembling a tree structure. In such tree-like topologies, the QR-MO agent can quickly learn the optimal path due to fewer choices at each node.

On the other hand, for 25N50E and 100N150E, the DPS reaches $10^{-2}$ and near 0, respectively, after 100 episodes. This means that the QR-MO achieves a near-optimal solution. These networks have high average node degrees of 4.0 and 3.0 edges per node, resembling mesh topologies. The QR-MO agent requires more episodes in such high-degree networks to identify efficient paths. Nevertheless, the decrease in DPS from $10^{-1}$ to near zero within 80 episodes for both networks highlights a good potential of RL algorithms also for mesh networks. For the 100N150E with a moderate average node degree, the QR-MO can return a near-optimal solution with DPS $=10^{-18}$.

Fig~\ref{fig:correctness}(a) illustrates the average correctness at episodes 10, 20, 50 and 100. Fig~\ref{fig:correctness}(b) depicts the average number of QR-MO solutions within the Pareto set. The higher episode improves the correctness and average number of correct solutions. In the 50N50E network, QR-MO has the best performance, reaching 100\% correctness in 20 episodes, followed by the MCC network, which reaches 100\% correctness in 50 episodes. QR-MO performs better in the 50N50E network than in the slightly denser MCC network, with the lowest average node degree of 2 edges per node. Nonetheless, after 50 episodes, all QR-MO solutions from both networks are correct. 

Regarding the correctness metric, QR-MO produces near-optimal correctness for the 25N50E and 100N150E networks in 100 episodes. Both networks show an increasing trend in correctness with more episodes. In the 25N50E network, QR-MO initially achieves an average correctness higher than the one in the 100N150E network, around 60\% instead of 20\% correctness at episode 10. Anyway, when the number of episodes increases, QR-MO performs better in the 100N150E network (100\% correctness and an average number of correct solutions of around 2.8 at episode 100) than in the 25N50E network (88\% correctness at episode 100).
The 40\% gap difference in correctness at 10 episodes between 25N50E and 100N150E is due to the total number of nodes difference, with the 100N150E having four times more nodes. However, at episode 100, the correctness of QR-MO in 100N150E is 10\%  better than that in 25N50E. This indicates that the number of nodes significantly affects the average correctness, particularly with fewer episodes. With more episodes (50,100), the performance difference is more impacted by the average node degree, and the lower average node degree network can outperform the higher one. 


The results presented in Fig.~\ref{fig:DPS-all} and Fig.~\ref{fig:correctness} indicate that the RL algorithm has promising potential for solving the MOSP problem and achieving near-optimal solutions in both small and large networks. By episode 100, the QR-MO algorithm had found near-optimal solutions across different networks. 
Anyway, QR-MO's processing time is in the order of thousands of milliseconds, whereas MDA generates solutions in hundreds of milliseconds. 
However, this refers to a static network condition, i.e., the MOSP problem is solved for a given combination of cost values, nodes, and edges.
Future works are needed to prove the profitability of QR-MO as highlighted in the next section.

%% file: 05_Conclusion.tex
\section{Conclusion and Future Works}
\label{sec:conclusion}

This paper explores the benefit of solving the MOSP problem in 5G-MEC systems by proposing QR-MO. QR-MO is an RL-based algorithm that modifies the classical Q-routing to accommodate multiple objectives and uses a heuristic procedure for selecting the action selection and storing the solutions. The QR-MO performance has been evaluated by comparing the solutions with the optimal solutions provided by MDA. We have introduced two performance metrics to evaluate the QR-MO performance: DPS and correctness. Four networks with different numbers of nodes and different average node degrees have been considered in the evaluation. In the case of tree-like networks with a small average node degree, QR-MO performs best in terms of DPS and correctness: fewer episodes can return correct solutions. The QR-MO also performs well in the case of a mesh network, but it can return correct solutions after a higher number of episodes. However, our work still has some limitations and challenges. 
Although the QR-MO has promising results, the processing time to reach the correct solution is hundreds to thousands of milliseconds. Meanwhile, MDA can generate solutions in tens to hundreds of milliseconds. Anyway, our evaluation is with static network conditions, further works need to be done to evaluate the profitability of QR-MO in dynamic networks. 
In dynamic network conditions, MDA must recompute the solutions from scratch at every network change.
QR-MO  can instead exploit the previous training to compute the solutions in the new network conditions. 
In this case, after convergence, QR-MO computes solutions for each episode, even when network conditions differ from those used during training and previous episodes. The QR-MO can implement its learned policy directly, generating solutions in about 5 ms on average. This suggests that QR-MO has the potential to adapt to dynamic changes in the network and adjust its policy in near real-time. In contrast, the MDA, which depends on static network assumptions, must generate solutions from scratch, requiring approximately 50 ms.